# Digital welfare fraud detection and the Dutch *SyRI* judgment

**Marvin van Bekkum**
Radboud University, Nijmegen, Gelderland, Netherlands

**Frederik Zuiderveen Borgesius**
Radboud University, Nijmegen, Gelderland, Netherlands

## Abstract
In 2020, a Dutch court passed judgment in a case about a digital welfare fraud detection system called *Systeem Risico Indicatie* (SyRI). The court ruled that the SyRI legislation is unlawful because it does not comply with the right to privacy under the European Convention of Human Rights. In this article we analyse the judgment and its implications. This ruling is one of first in which a court has invalidated a welfare fraud detection system for breaching the right to privacy. We show that the immediate effects of the judgment are limited. The judgment does not say much about automated fraud detection systems in general, because it is limited to the circumstances of the case. Still, the judgment is important. The judgment reminds policymakers that fraud detection must happen in a way that respects data protection principles and the right to privacy. The judgment also confirms the importance of transparency if personal data are used.

## Introduction

In February 2020, the Court at First Instance in The Hague, the Netherlands, ruled in the *SyRI* case, about an automated welfare fraud detection system called SyRI.[1] The court decided that the SyRI

---

1. NJCM et al. v The Dutch State (2020) The Hague District Court ECLI: NL: RBDHA:2020:1878 (*SyRI*). English translation Available at: http://deeplink.rechtspraak.nl/uitspraak?id=ECLI:NL:RBDHA:2020:1878.

---

**Corresponding author:**
Frederik Zuiderveen Borgesius, Professor ICT and law at the iHub and the Institute for Computing and Information Sciences (iCIS), Radboud University, Houtlaan 4, Nijmegen, 6525 XZ, Netherlands.
E-mail: frederikzb@cs.ru.nl

legislation did not strike a fair balance between the interest in fraud detection on the one hand, and the human right to privacy on the other. Therefore, the court declared that SyRI is unlawful. As the UN Special Rapporteur on extreme poverty and human rights noted, this is one of the first cases in which a court anywhere in the world stopped the use of digital welfare technologies on human rights grounds.[2]

In this article we analyse the judgment and its implications. We discuss the judgment in such a way that lawyers who are not specialised in Dutch and European law can follow the discussion. The following questions lead the discussion in this article. What was the Dutch SyRI system? What are the main points of the *SyRI* judgment? What are the implications, if any, of the *SyRI* judgment for state-run fraud detection systems and automated decision systems?

The article's contributions to scholarship are the following. This is one of the most in-depth scholarly discussions of the judgment to date. In addition, the article describes what SyRI is in more detail than other publications, based on an examination of the legislative history of SyRI. Moreover, there has not been much academic writing on the judgment in English (other scholarly publications about the judgment in English are: Meuwese, 2020; Bekker, 2021; Wieringa, 2021). There were academic publications in Dutch about the judgment; we refer to those publications throughout the article.

The article is structured in sections as follows. First, we contextualise the *SyRI* judgment by making some remarks about the digitalisation of the welfare state. Next, we explain what SyRI was and its background. The next section introduces the group that sued the state and summarises the judgment. In the following section, we analyse the judgment, discuss the implications of the judgment for fraud detection and automated decision systems, and highlight some developments after the judgment. The final section concludes.

## The digitisation of the welfare state can erode human rights

The Dutch SyRI legislation fits in with a worldwide trend in digitisation of the welfare state. The Covid-19 virus has caused a financial crisis, causing more people to depend on welfare. In many countries, governments aim to use new technology to make the welfare state more efficient and to mitigate welfare fraud. However, this digitisation of the welfare state can also erode people's rights.

The U.N. Special Rapporteur on extreme poverty and human rights, Philip Alston, has expressed his concern about the digital welfare state on several occasions. The Special Rapporteur made a country visit to the United Kingdom in 2018. He noted that the 'welfare state is gradually disappearing behind a webpage and an algorithm, with significant implications for those living in poverty' (U.N. Human Rights Council, 2019: 13). According to the Special Rapporteur, more transparency about how fraud detection systems work is required. He noted that '[i]n the absence of transparency about the existence and workings of automated systems, the rights to contest an adverse decision and to seek a meaningful remedy are illusory' (U.N. Human Rights Council, 2019: 15).

---

2. U.N. Special Rapporteur on extreme poverty and human rights (2020) Landmark ruling by Dutch court stops government attempts to spy on the poor – UN expert. Available at: https://www.ohchr.org/EN/NewsEvents/Pages/DisplayNews.aspx?LangID=E&NewsID=25522 (accessed 16 June 2021).

In his latest report about digital welfare states worldwide, the Special Rapporteur concluded that states should search for technologies that improve welfare instead of technologies that aim to detect fraud (U.N. Human Rights Council, 2019: 23). He stated that significant changes should be made to avoid 'stumbling, zombie-like, into a digital welfare dystopia. Such a future would be one in which unrestricted data-matching is used to expose and punish the slightest irregularities in the record of welfare beneficiaries ( . . . )'(U.N. Human Rights Council, 2019: 21–22). As the rest of this article shows, the Dutch SyRI system is a good example of such unrestricted data-matching.

## The SyRI system

### What is SyRI?

Over the past 20 years, politicians in the Netherlands have focused a lot on fighting welfare fraud (Parlementaire ondervragingscommissie Kinderopvangtoeslag, 2021: 1). The history of SyRI starts in 2003. Several Dutch administrative organs agreed to collaborate and to exchange data,[3] using the SyRI system, to reduce fraud.[4] After a change in legislation in 2013, more administrative organs were allowed to use SyRI.[5] The Dutch Data Protection Authority (Dutch Data Protection Authority (College Bescherming Persoonsgegevens), 2014) and the Dutch Council of State (which provides advice on legal Bills) (Raad van State (Council of State), 2014) both criticised the proposed SyRI system, because the proposals hardly limited the amount of personal data that could be collected. The Dutch government largely ignored the advice of these two bodies.[6] The Dutch Parliament adopted the SyRI legislation without debate (for more details about the SyRI history, see Gantchev, 2019: 17; Jansen and Reijneveld, 2020; Wieringa, 2021).

The SyRI legislation does not concretely define SyRI. Only the government's explanatory memorandum of the SyRI legislation includes the following, rather open, description:

> SyRI includes the technical infrastructure and associated procedures through which data can be linked and thereafter anonymously analysed in a secure environment, in order to generate risk notifications. A risk notification means that a legal person or natural person is considered worth investigating in connection with possible fraud, unlawful use of and noncompliance with legislation. This method leads to a more effective and efficient use of the control instrument.[7]

Hence, SyRI is a socio-technical infrastructure. It consists of technology combined with several procedures for generating risk notifications. These risk notifications are sent to the administrative bodies that participate in the use of SyRI.[8]

---

3. The relevant administrative organs are: the Dutch tax authority, the Ministry of Social Affairs and Employment, the police, the public prosecutor's office, several administrative welfare bodies, and a number of municipalities.
4. Explanatory memorandum to the Dutch Wet SUWI: 2-3.
5. Wet structuur uitvoeringsorganisatie werk en inkomen (SUWI) [Work and Income (Implementation Organisation Structure Act], Art. 64.
6. De Minister van Sociale Zaken en Werkgelegenheid [Minister of Social Affairs and Employment] (2014) Nota van toelichting. Staatscourant 2014, 26306. Available at: https://zoek.officielebekendmakingen.nl/stcrt-2014-26306.html.
7. Explanatory memorandum to the Dutch Wet SUWI: 3. Translated from Dutch by the authors.
8. In some cases, risk notifications are also sent to other bodies than those that participated.

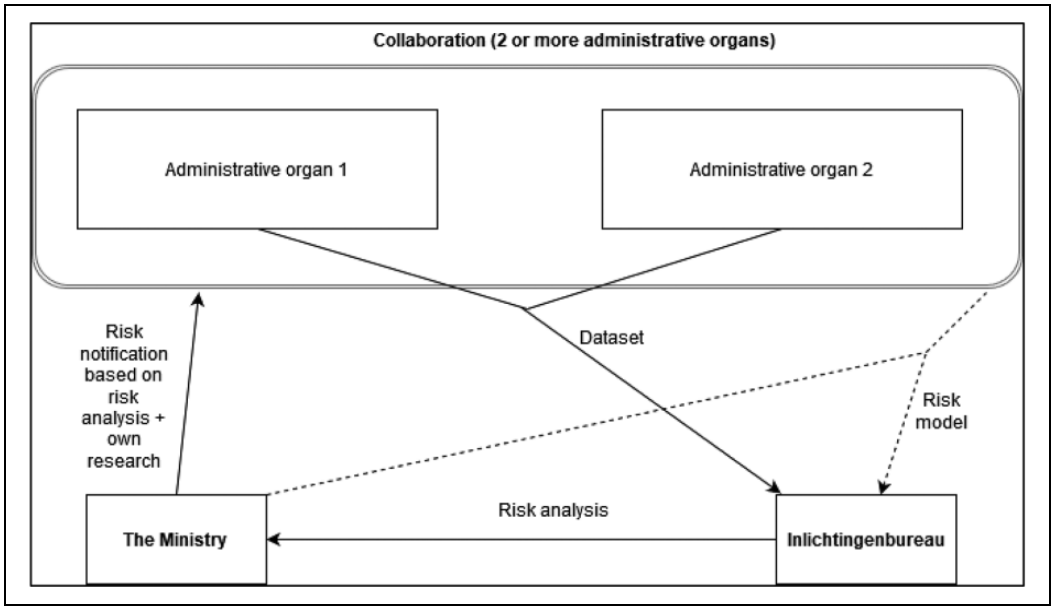

**Figure 1.** The data flows in SyRI projects.

## The data flows of SyRI projects

The data flows of SyRI projects are as follows. Participating administrative organs, in collaboration with the Ministry of Welfare and Social Affairs (hereafter 'the Ministry'), first determine the risk model to be used for fraud detection. Participating organs provide the data to a foundation known as the Inlichtingenbureau, which can be translated as 'Information Bureau'.[9]

The Information Bureau then pseudonymises the data from the administrative organs, combines the data and then analyses the differences in the dataset using the risk model. Pseudonymisation means, in short, deleting a person's name from a file and replacing the name with a number or other unique identifier. Suspicious cases are de-pseudonymised.

The process creates a risk analysis, which essentially contains the de-pseudonymised information about only the cases that are deemed suspicious by the algorithm.[10] The risk analysis is sent to the Ministry. The Ministry then conducts its own research on the dataset, based on the risk analysis it has received from the Information Bureau. We summarise the dataflows in SyRI projects in Figure 1.

(As an aside: after pseudonymisation, personal data typically remain personal data, because they can be linked back to the relevant individual.[11] Pseudonymisation is thus not a type of anonymisation. Anonymisation should ensure that data cannot be linked to an individual at all.)[12]

---

9. Inlichtingenbureau (2020) Over ons [about us]. https://www.inlichtingenbureau.nl/Over-ons (accessed 16 June 2021).
10. See section "Technical aspects of SyRI" for the known details about the analysis process.
11. See also the definition of pseudonymous data in General Data Protection Regulation, Art. 4(5).
12. See, for more detail on pseudonymous data, Zuiderveen Borgesius, 2016.

## Technical aspects of SyRI

The explanatory memorandum shows that SyRI projects are executed in two phases. In this section, we describe the risk model and the comparison algorithm used in SyRI projects (phase 1) and the investigation by the Ministry (phase 2).

In the first phase, the Information Bureau applies the risk model to the dataset. The Information Bureau first alters the dataset. Identifiable information such as names, citizen service numbers and addresses are replaced by a code. The data are thus pseudonymised.[13] That code is generated using a key that is only known to the Information Bureau.[14] The resulting dataset is called the 'source file'. The Information Bureau also keeps a separate 'key file' containing the codes and the corresponding identifiable information that was replaced.

The Information Bureau compares the source file with the risk indicators from the predefined risk model, using an algorithm. There is no public information available about the algorithm used. The output of this comparison is a list containing the pseudonym codes of suspicious cases. This list of codes is then used together with the key file to de-pseudonymise parts of the source file. All the information about suspicious cases that are 'associated with increased risk' is sent to the Ministry, except for the key file itself.[15]

The concrete risk model used in practical applications of SyRI remains secret. The legislator gives a few examples of risk indicators for the risk models.

- If two people live together, but report to the state that they live at different addresses, they probably claim too much in welfare benefits (a couple that lives separately can claim more benefits in the Netherlands).
- A person has probably concealed assets if the balance of his or her bank account increases sharply within a year.
- A person has probably concealed assets if he or she has multiple garages and has owned multiple vehicles.

In the second phase, the Ministry examines the information about the suspicious cases. The exact examination method used is kept secret. The examination concerns drawing 'relational connections *between* potential hits'.[16]

## Lacuna in knowledge about SyRI

Three components of SyRI are kept strictly secret in the government's explanatory memorandum. First, there is no public information available about the exact risk model that is used for the comparison. Second, the algorithm that makes the comparison is secret. Third, the government does not provide information about the method used by the Ministry to find connections between hits.[17]

---

13. The explanatory memorandum claims that the dataset is 'encrypted', but the dataset is merely pseudonymised. See Explanatory memorandum to the Dutch Wet SUWI: 13, where the terms are intermixed.
14. Explanatory memorandum to the Dutch Wet SUWI: 5 and 38.
15. Explanatory memorandum to the Dutch Wet SUWI: 14.
16. Explanatory memorandum to the Dutch Wet SUWI: 5 and 38.
17. The court in the SyRI case noticed the same problem. The state did not give the judges more information about the inner workings of these three components than what follows from the explanatory memorandum.

It is not clear from the legislation or the explanatory memorandum whether SyRI is one centralised system or consists of several different, decentralised systems. The legislator mentions an 'infrastructure'. That word seems to imply that SyRI can consist of multiple systems. Finally, it is unclear from the explanatory memorandum whether the technical infrastructure also includes the analysis performed by the Ministry in the second phase of a SyRI project.

The government also refused to give the information to the Dutch Parliament, out of fear that the information would become public. The government argued that potential fraudsters could use that information in order to escape the algorithms' risk analyses (Letter of the State Secretary, 2017: 1–2). Indeed, we will see below that the court also noted a lack of transparency regarding SyRI.

## The *SyRI* judgment: the main points

### The group that sued the state

Before we summarise the main points of the judgment,[18] we first provide some background about the group that sued the state. Six NGOs formed a group to sue the state over SyRI:

i. The Dutch Section of the International Commission of Jurists (NJCM). The NJCM works to promote and protect human rights in the Netherlands.[19]
ii. Platform Bescherming Burgerrechten (Civil Rights Protection Platform), an NGO that focuses on the protection of traditional civil rights.
iii. Privacy First, a Dutch foundation with the aim to preserve and promote the right to privacy.[20]
iv. Stichting KDVP, a foundation committed to maintaining privacy and confidentiality within the (mental) health care sector.[21]
v. De Landelijke Clientenraad (National Client Participation Council). This organisation aims to fulfil a key role pertaining to client participation regarding policy issues in the area of work and income, among other things.[22]
vi. Federatie Nederlandse Vakbeweging, or FNV (the Netherlands Trade Union Confederation) is a large trade union. The FNV promotes the interests of its members, guided by – among other things – the fundamental values of equality of all persons, freedom, justice and solidarity.[23]

---

18. We do not discuss all aspects of the judgment. For instance, we do not discuss standing and procedural law. We quote from the English translation of the judgment: NJCM et al. v The Dutch State (2020) The Hague District Court ECLI: NL: RBDHA:2020:1878 (*SyRI*). Available at: https://uitspraken.rechtspraak.nl/inziendocument?id=ECLI:NL:RBDHA:2020:1878.
19. Juristen Comité voor Mensenrechten, https://njcm.nl.
20. Privacy First, https://www.privacyfirst.eu.
21. Stichting KDVP, https://www.kdvp.nl/over-kdvp/doelstellingen.
22. Landelijke Cliëntenraad, http://www.landelijkeclientenraad.nl/Wie-zijn-wij-en-wat-doen-wij.
23. FNV, https://www.fnv.nl. For procedural reasons that are outside the scope of this article, the FNV was listed in the judgment as a separate party, rather than as a part of the group. See NJCM et al. v The Dutch State (2020) The Hague District Court ECLI: NL: RBDHA:2020:1878 (*SyRI*). Available at: https://uitspraken.rechtspraak.nl/inziendocument?id=ECLI:NL:RBDHA:2020:1878 [2].

Lastly, two famous authors joined the group: Tommy Wieringa and Maxim Februari. Wieringa has written a number of successful novels and other books and writes regular columns.[24] Februari is a philosopher and a writer, and is also a columnist for a large newspaper.[25]

The group also started a campaign called 'Bij Voorbaat Verdacht', which can be translated as 'Suspected in Advance'.[26] The campaign aimed, among other things, to raise awareness and trigger public discussion about the SyRI system. The two writers and various members of the group often spoke about the *SyRI* case in the media, including on popular TV talk shows.

We use the phrase 'NJCM et al.' to refer to all the parties that sued the state.[27]

One last party deserves to be mentioned: UN special rapporteur Philip Alston (previously mentioned in the second section of this article, where we placed the judgment in the worldwide trend of the digital welfare state). Alston submitted an Amicus Brief to the Court in the *SyRI* case. The letter discussed the SyRI system in the context of the digital welfare state. Alston was not one of the parties that sued the state, but his letter was probably influential for the outcome of the case (for more information about the Amicus Brief see Gantchev, 2021: 4.3). It is unusual in Dutch court cases for an international organisation or a UN official to submit an Amicus Brief.

## Interference with the right to privacy

Roughly summarised, NJCM et al. asked the court to rule that parts of the SyRI legislation were incompatible with the right to privacy, as protected in Article 8 of the European Convention on Human Rights and in other human rights treaties.[28]

The court tested the SyRI legislation against the right to private life in Article 8 European Convention on Human Rights (ECHR).[29] Article 8 grants everyone the right to respect for their private life – privacy, for short.[30] Referring to case law by the European Court of Human Rights, the Dutch court stated that the right to private life has a broad scope.[31] The court used the GDPR's data protection principles to interpret Article 8 ECHR: '[t]he court will take into account the ( . . . ) general principles of data protection from the Charter [of Fundamental Rights of the European Union] and the GDPR in its consideration of whether the SyRI legislation meets the requirements of Article 8 ECHR.'[32]

---

24. Tommy Wieringa, https://www.debezigebij.nl/auteur/tommy-wieringa/. We particularly recommend his novel 'Joe Speedboat'. For procedural reasons that are outside the scope of this article, the two famous authors were declared inadmissible.
25. Maxim Februari, https://www.nrc.nl/rubriek/maxim-februari/. For procedural reasons that are outside the scope of this article, the two famous authors were declared inadmissible.
26. Bij voorbaat verdacht, https://bijvoorbaatverdacht.nl/over/.
27. NJCM, https://njcm.nl. Zuiderveen Borgesius, one of the authors of this article, has a connection with the NJCM. He is a member of the editorial board of the NCJM's journal, Nederlands Tijdschrift voor Mensenrechten (Dutch human rights journal).
28. NJCM et al. v The Dutch State (2020) The Hague District Court ECLI: NL: RBDHA:2020:1878 (*SyRI*). Available at: https://uitspraken.rechtspraak.nl/inziendocument?id=ECLI:NL:RBDHA:2020:1878 [5.1]. NJCM et al. brought other claims, too, but the court dismissed those claims. See para. 6.113 and further.
29. NJCM et al. v The Dutch State (2020) The Hague District Court ECLI: NL: RBDHA:2020:1878 (*SyRI*). Available at: https://uitspraken.rechtspraak.nl/inziendocument?id=ECLI:NL:RBDHA:2020:1878 [6.20-6.26].
30. NJCM et al. v The Dutch State (2020) The Hague District Court ECLI: NL: RBDHA:2020:1878 (*SyRI*). Available at: https://uitspraken.rechtspraak.nl/inziendocument?id=ECLI:NL:RBDHA:2020:1878 [6.21]. See on the (small) difference between 'privacy' and 'private life' González Fuster, 2014: 82–84; 255.
31. NJCM et al. v The Dutch State (2020) The Hague District Court ECLI: NL: RBDHA:2020:1878 (*SyRI*). Available at: https://uitspraken.rechtspraak.nl/inziendocument?id=ECLI:NL:RBDHA:2020:1878 [6.24]. From the right to a private life, the court even deduced a right to fair procedures and a right to non-discrimination: 'the right to respect for private life in the context of data processing concerns the right to equal treatment in equal cases, and the right to protection against discrimination, stereotyping and stigmatisation.'

The parties (NJCM et al. and the state) agreed that the SyRI legislation interfered with the right to private life.[33] The question was whether 'the SyRI legislation meets the requirements of Article 8 paragraph 2 ECHR to justify that interference.'[34] The court noted the three criteria for justifying an interference: (1) in accordance with the law, (2) necessary in a democratic society, and (3) for a legitimate purpose.[35]

The interference has a legitimate purpose. The court noted that 'social security is one of the pillars of Dutch society and contributes to a considerable extent to prosperity in the Netherlands.' The court added that combating fraud is important, and that it makes sense that the state uses new technologies to combat fraud.[36]

In the *SyRI* case, the court found that the application of the criterion 'prescribed by law' is closely intertwined with the criterion 'necessary in a democratic society'.[37] We will therefore skip the discussion of the 'prescribed by law' criterion. In the following sections, we first discuss the extent and seriousness of the interference. We then discuss the application of the criterion 'necessary in a democratic society'.

## Extent and seriousness of the interference: general

The court first discussed the extent to which SyRI interfered with the right to private life. The court concluded that 'the amount of data that can be used in the application of SyRI is substantial. A total of 17 data categories of various types qualify. Each separate category can be deemed to encompass a large amount of data. Depending on the specific SyRI project, there may be large amounts of structured data sets from various sources.'[38]

The court mentioned the uncertainty regarding what SyRI is.[39] NJCM et al argued that SyRI constituted a 'dragnet, untargeted approach in which personal data are collected for investigation purposes.'[40] The state, on the other hand, asserted that 'only data from existing data sets of designated government or other bodies are compared.'[41]

---

32. NJCM et al. v The Dutch State (2020) The Hague District Court ECLI: NL: RBDHA:2020:1878 (*SyRI*). Available at: https://uitspraken.rechtspraak.nl/inziendocument?id=ECLI:NL:RBDHA:2020:1878 [6.41].
33. NJCM et al. v The Dutch State (2020) The Hague District Court ECLI: NL: RBDHA:2020:1878 (*SyRI*). Available at: https://uitspraken.rechtspraak.nl/inziendocument?id=ECLI:NL:RBDHA:2020:1878 [6.42].
34. NJCM et al. v The Dutch State (2020) The Hague District Court ECLI: NL: RBDHA:2020:1878 (*SyRI*). Available at: https://uitspraken.rechtspraak.nl/inziendocument?id=ECLI:NL:RBDHA:2020:1878 [6.43].
35. NJCM et al. v The Dutch State (2020) The Hague District Court ECLI: NL: RBDHA:2020:1878 (*SyRI*). Available at: https://uitspraken.rechtspraak.nl/inziendocument?id=ECLI:NL:RBDHA:2020:1878 [6.21].
36. NJCM et al. v The Dutch State (2020) The Hague District Court ECLI: NL: RBDHA:2020:1878 (*SyRI*). Available at: https://uitspraken.rechtspraak.nl/inziendocument?id=ECLI:NL:RBDHA:2020:1878 [6.3].
37. NJCM et al. v The Dutch State (2020) The Hague District Court ECLI: NL: RBDHA:2020:1878 (*SyRI*). Available at: https://uitspraken.rechtspraak.nl/inziendocument?id=ECLI:NL:RBDHA:2020:1878 [6.71-6.72].
38. NJCM et al. v The Dutch State (2020) The Hague District Court ECLI: NL: RBDHA:2020:1878 (*SyRI*). Available at: https://uitspraken.rechtspraak.nl/inziendocument?id=ECLI:NL:RBDHA:2020:1878 [6.50].
39. NJCM et al. v The Dutch State (2020) The Hague District Court ECLI: NL: RBDHA:2020:1878 (*SyRI*). Available at: https://uitspraken.rechtspraak.nl/inziendocument?id=ECLI:NL:RBDHA:2020:1878 [6.44].
40. NJCM et al. v The Dutch State (2020) The Hague District Court ECLI: NL: RBDHA:2020:1878 (*SyRI*). Available at: https://uitspraken.rechtspraak.nl/inziendocument?id=ECLI:NL:RBDHA:2020:1878 [6.45].
41. NJCM et al. v The Dutch State (2020) The Hague District Court ECLI: NL: RBDHA:2020:1878 (*SyRI*). Available at: https://uitspraken.rechtspraak.nl/inziendocument?id=ECLI:NL:RBDHA:2020:1878 [6.47].

As previously mentioned,[42] neither the statement by NJCM nor the statement by the State could be proven from the SyRI legislation. The court similarly concluded that '[t]here currently are no indications of "deep learning" or data mining or the development of risk profiles in the implementation of the SyRI legislation.' However, the court, noted that 'the SyRI legislation does provide scope for the development and application of a risk model using "deep learning" and data mining, and for the development of risk profiles.'[43]

The court indicated a lack of transparency regarding SyRI.[44] The risk model and its indicators are not public nor known to the data subjects. Additionally, the legislation contains no duty to inform data subjects that their data has been processed or that a risk report has been submitted.[45]

## Extent and seriousness of the interference: profiling and automated individual decision-making?

NJCM et al. also invoked Article 22 of the GDPR, which contains an in-principle prohibition of fully automated decisions with legal or similarly significant effects (see, for analysis of Article 22 of the GDPR, Mendoza and Bygrave, 2017). Violating such a prohibition would make the interference to the right to privacy more serious. NJCM et al. argued 'that the submission of a risk report (...) can be considered a decision with legal effect, or at least a decision that affects the data subjects significantly in another way (...) within the meaning of Article 22 GDPR.'[46]

The court agreed that the submission of a risk report had a significant effect in the sense of Article 22 GDPR. However, the court noted that submitting such a risk report did not have legal effect.[47] The court did not 'give an opinion on whether the exact definition of automated individual decision-making in the GDPR and, insofar as this is the case, one or more of the exceptions to the prohibition in the GDPR have been met. That is irrelevant in the context of the review by the court whether the SyRI legislation meets the requirements of Article 8 ECHR.'[48]

Irrespective of whether SyRI projects apply fully automated decision-making, however, the court noted that the significant effect of the submission of a risk report and its inclusion into risk registers was a relevant factor for the assessment of whether the SyRI legislation violated Article 8 ECHR.[49]

---

42. See section "Lacuna in knowledge about SyRI" of this article.
43. NJCM et al. v The Dutch State (2020) The Hague District Court ECLI: NL: RBDHA:2020:1878 (*SyRI*). Available at: https://uitspraken.rechtspraak.nl/inziendocument?id=ECLI:NL:RBDHA:2020:1878 [6.63]. See also para. 6.47, 6.49, 6.51.
44. NJCM et al. v The Dutch State (2020) The Hague District Court ECLI: NL: RBDHA:2020:1878 (*SyRI*). Available at: https://uitspraken.rechtspraak.nl/inziendocument?id=ECLI:NL:RBDHA:2020:1878 [6.65].
45. NJCM et al. v The Dutch State (2020) The Hague District Court ECLI: NL: RBDHA:2020:1878 (*SyRI*). Available at: https://uitspraken.rechtspraak.nl/inziendocument?id=ECLI:NL:RBDHA:2020:1878 [6.65].
46. NJCM et al. v The Dutch State (2020) The Hague District Court ECLI: NL: RBDHA:2020:1878 (*SyRI*). Available at: https://uitspraken.rechtspraak.nl/inziendocument?id=ECLI:NL:RBDHA:2020:1878 [5.57].
47. NJCM et al. v The Dutch State (2020) The Hague District Court ECLI: NL: RBDHA:2020:1878 (*SyRI*). Available at: https://uitspraken.rechtspraak.nl/inziendocument?id=ECLI:NL:RBDHA:2020:1878 [6.59].
48. NJCM et al. v The Dutch State (2020) The Hague District Court ECLI: NL: RBDHA:2020:1878 (*SyRI*). Available at: https://uitspraken.rechtspraak.nl/inziendocument?id=ECLI:NL:RBDHA:2020:1878 [6.60].
49. NJCM et al. v The Dutch State (2020) The Hague District Court ECLI: NL: RBDHA:2020:1878 (*SyRI*). Available at: https://uitspraken.rechtspraak.nl/inziendocument?id=ECLI:NL:RBDHA:2020:1878 [6.60].

### Necessary in a democratic society: proportionality and subsidiarity

For the court, the main question was 'whether the SyRI legislation meets the requirements of necessity, proportionality and subsidiarity pursuant to Article 8 paragraph 2 ECHR in light of the aims it pursues. There has to be a "fair balance" between the purposes of the SyRI legislation and the invasion of private life the legislation causes.'[50]

Referring to the *Marper* judgment of the European Court of Human Rights, the Dutch court found that the State has a special responsibility for striking the right balance between the benefits of applying new technologies on the one hand, and the potential interference with the right to privacy on the other hand.[51]

In its discussion of the criterion 'necessary in a democratic society', the court mentioned several data protection principles from the GDPR: the transparency, purpose limitation and data minimisation principles.

*Lack of transparency.* The court noted that '[t]he right to respect for private life also means that a data subject must reasonably be able to track their personal data.'[52] The court found that 'the SyRI legislation is insufficiently transparent and verifiable to conclude that the interference with the right to respect for private life which the use of SyRI may entail is necessary, proportional and proportionate in relation to the aims the legislation pursues.'[53]

The court added that 'the SyRI legislation in no way provides information on (...) which objective factual data can justifiably lead to the conclusion that there is an increased risk.'[54] Moreover, as mentioned,[55] the SyRI legislation is opaque in respect of the risk model, the type of algorithm used in the model, and the risk analysis method.[56]

*Discrimination risks.* According to the court, discrimination is particularly important, because SyRI brings risks of discrimination. It asserted: 'there is in fact a risk that SyRI inadvertently creates links based on bias, such as a lower socio-economic status or an immigration background'. Moreover, according to the court, to date, SyRI has only been applied in poor neighbourhoods ('problem districts').[57]

---

50. NJCM et al. v The Dutch State (2020) The Hague District Court ECLI: NL: RBDHA:2020:1878 (*SyRI*). Available at: https://uitspraken.rechtspraak.nl/inziendocument?id=ECLI:NL:RBDHA:2020:1878 [6.80].
51. NJCM et al. v The Dutch State (2020) The Hague District Court ECLI: NL: RBDHA:2020:1878 (*SyRI*). Available at: https://uitspraken.rechtspraak.nl/inziendocument?id=ECLI:NL:RBDHA:2020:1878 [6.6and6.84]. See S. and Marper v the United Kingdom (2008) ECtHR 30562/04 and 30566/04.
52. NJCM et al. v The Dutch State (2020) The Hague District Court ECLI: NL: RBDHA:2020:1878 (*SyRI*). Available at: https://uitspraken.rechtspraak.nl/inziendocument?id=ECLI:NL:RBDHA:2020:1878 [6.90].
53. NJCM et al. v The Dutch State (2020) The Hague District Court ECLI: NL: RBDHA:2020:1878 (*SyRI*). Available at: https://uitspraken.rechtspraak.nl/inziendocument?id=ECLI:NL:RBDHA:2020:1878 [6.86].
54. NJCM et al. v The Dutch State (2020) The Hague District Court ECLI: NL: RBDHA:2020:1878 (*SyRI*). Available at: https://uitspraken.rechtspraak.nl/inziendocument?id=ECLI:NL:RBDHA:2020:1878 [6.87].
55. Section "Lacuna in knowledge about SyRI" of this article.
56. NJCM et al. v The Dutch State (2020) The Hague District Court ECLI: NL: RBDHA:2020:1878 (*SyRI*). Available at: https://uitspraken.rechtspraak.nl/inziendocument?id=ECLI:NL:RBDHA:2020:1878 [6.89].
57. NJCM et al. v The Dutch State (2020) The Hague District Court ECLI: NL: RBDHA:2020:1878 (*SyRI*). Available at: https://uitspraken.rechtspraak.nl/inziendocument?id=ECLI:NL:RBDHA:2020:1878 [6.93].

The court added that 'the right to respect for private life in the context of data processing concerns the right to equal treatment in equal cases, and the right to protection against discrimination, stereotyping and stigmatisation.'[58]

*Purpose limitation principle and the data minimisation principle.* The court also concluded that SyRI breached the GDPR's purpose limitation and data minimisation principles. The principle of purpose limitation means that personal data must be collected for specified, explicit and legitimate purposes and not further processed in a manner that is incompatible with those purposes.[59] The data minimisation principle requires personal data to be adequate, relevant and limited to what is necessary in relation to the purposes for which they are processed.[60]

However, the court observed that under the SyRI legislation, almost any type of personal data could be used for analysis.[61] In light of the data protection principles, the Court therefore found insufficient safeguards for compliance with Article 8(2) ECHR.[62]

*Data Protection Impact Assessment.* In addition, the court discussed a concept introduced by the GDPR: a Data Protection Impact Assessment (DPIA).[63] A DPIA can be described as 'a process designed to help [an organisation] systematically analyse, identify and minimise the data protection risks of a project or plan.'[64]

The court noted that the state did not conduct DPIAs for the separate SyRI projects. The state argued that it carried out a DPIA of the SyRI legalisation as a whole, and that therefore DPIAs were not needed for the separate SyRI projects.[65] That argument did not convince the court. First, the state's DPIA was conducted before the GDPR applied. Second, the state failed to explain why a DPIA was not needed for each separate SyRI project.[66]

---

58. NJCM et al. v The Dutch State (2020) The Hague District Court ECLI: NL: RBDHA:2020:1878 (*SyRI*). Available at: https://uitspraken.rechtspraak.nl/inziendocument?id=ECLI:NL:RBDHA:2020:1878 [6.24].
59. NJCM et al. v The Dutch State (2020) The Hague District Court ECLI: NL: RBDHA:2020:1878 (*SyRI*). Available at: https://uitspraken.rechtspraak.nl/inziendocument?id=ECLI:NL:RBDHA:2020:1878 [6.32]; General Data Protection Regulation Art. 5(1)(b).
60. NJCM et al. v The Dutch State (2020) The Hague District Court ECLI: NL: RBDHA:2020:1878 (*SyRI*). Available at: https://uitspraken.rechtspraak.nl/inziendocument?id=ECLI:NL:RBDHA:2020:1878 [6.33]; General Data Protection Regulation Art. 5(1)(c).
61. NJCM et al. v The Dutch State (2020) The Hague District Court ECLI: NL: RBDHA:2020:1878 (*SyRI*). Available at: https://uitspraken.rechtspraak.nl/inziendocument?id=ECLI:NL:RBDHA:2020:1878 [6.89].
62. NJCM et al. v The Dutch State (2020) The Hague District Court ECLI: NL: RBDHA:2020:1878 (*SyRI*). Available at: https://uitspraken.rechtspraak.nl/inziendocument?id=ECLI:NL:RBDHA:2020:1878 [6.106].
63. See General Data Protection Regulation Art. 35.
64. UK Information Commissioner's Office Data protection impact assessments. https://ico.org.uk/for-organisations/guide-to-data-protection/guide-to-the-general-data-protection-regulation-gdpr/accountability-and-governance/data-protection-impact-assessments/ (accessed 16 June 2021).
65. NJCM et al. v The Dutch State (2020) The Hague District Court ECLI: NL: RBDHA:2020:1878 (*SyRI*). Available at: https://uitspraken.rechtspraak.nl/inziendocument?id=ECLI:NL:RBDHA:2020:1878 [6.103].
66. NJCM et al. v The Dutch State (2020) The Hague District Court ECLI: NL: RBDHA:2020:1878 (*SyRI*). Available at: https://uitspraken.rechtspraak.nl/inziendocument?id=ECLI:NL:RBDHA:2020:1878 [6.105]. The Court adds that only five SyRI projects appear to have been conducted.

*Conclusion.* Because the SyRI legislation is not 'necessary in a democratic society', the court found a violation of Article 8 ECHR.[67] Indeed, it asserted that 'the legislation does not strike the "fair balance" required under the ECHR between the social interest the legislation serves and the violation of private life to which the legislation gives rise'.[68]

## The *SyRI* judgment: analysis

With 36 pages, the SyRI judgment is long and detailed for a judgment by a lower court. The court also published an English translation, which is rare. Apparently, the court realised the judgment's importance and the international interest in it. Below, we comment on some particulars of the judgment.[69]

### Testing the SyRI legislation against human rights

As noted, the court assessed whether the SyRI legislation complies with the right to a private life, as protected by Article 8 ECHR. The ECHR is a treaty that was adopted in 1950 by the Council of Europe.[70]

Some readers may wonder why the court did not test the SyRI legislation against the Dutch Constitution, which also protects the right to private life.[71] The reason is that the Dutch Constitution prohibits judges from testing Acts of Parliament (legislation) against the Constitution[72] (Besselink, 2012: 5). In practice, however, the constitutional provision poses few problems. As the SyRI judgment illustrates, courts can test legislation against human rights based on human rights treaties.[73]

### The interplay between the GDPR and the right to a private life

As discussed, the court used the GDPR principles to interpret, or operationalise, the right to a private life under the ECHR. The court's strategy seems peculiar (Dommering, 2020). Why did the court not test the SyRI legislation directly against the GDPR?

---

67. NJCM et al. v The Dutch State (2020) The Hague District Court ECLI: NL: RBDHA:2020:1878 (*SyRI*). Available at: https://uitspraken.rechtspraak.nl/inziendocument?id=ECLI:NL:RBDHA:2020:1878 [6.7and6.72].
68. NJCM et al. v The Dutch State (2020) The Hague District Court ECLI: NL: RBDHA:2020:1878 (*SyRI*). Available at: https://uitspraken.rechtspraak.nl/inziendocument?id=ECLI:NL:RBDHA:2020:1878 [6.7].
69. We do not comment on all aspects of the judgment. For instance, we do not discuss standing and procedural law.
70. The most important human rights organisation in Europe, with 48 Member States. Council of Europe, https://www.coe.int (accessed 16 June 2021). The Council of Europe is a different organisation to the European Union (27 Member States). On the EU, see: European Union, https://europa.eu/european-union/index_en (accessed 16 June 2021).
71. Dutch Constitution Art. 10. https://www.government.nl/documents/regulations/2012/10/18/the-constitution-of-the-kingdom-of-the-netherlands-2008 accessed 16 June 2021.
72. Dutch Constitution Art. 120. Translation from https://www.servat.unibe.ch/icl/nl00000_.html (accessed 16 June 2021). The idea behind this prohibition of judicial review is that laws should be made by the democratically-elected parliament, and not by courts.
73. NJCM et al. v The Dutch State (2020) The Hague District Court ECLI: NL: RBDHA:2020:1878 (*SyRI*). Available at: https://uitspraken.rechtspraak.nl/inziendocument?id=ECLI:NL:RBDHA:2020:1878 [6.20]. See also Dutch constitution Art. 94: 'Statutory regulations in force within the Kingdom are not applicable if such application is in conflict with provisions of treaties or of resolutions by international institutions that are binding on all persons.' Translation from https://www.servat.unibe.ch/icl/nl00000_.html (accessed 16 June 2021).

We outline some possible explanations below. First, perhaps the court thought that testing a statute against a human rights treaty sounds more convincing. Invoking a human right makes a stronger impression than testing a statute against a detailed regulation such as the GDPR, which is full of legalese. Second, perhaps the court focused on the ECHR because NJCM et al. did so during the proceedings. The court noted that '[t]he focal point of the arguments of NJCM et al. is the alleged violation of Article 8 ECHR'.[74]

Third, perhaps the court wanted to prevent detailed discussions about the exact scope of the GDPR. For instance, police investigations fall outside the scope of the GDPR, and are governed by other EU data protection rules.[75] Therefore, certain fraud investigations might fall, at least partly, outside the scope of the GDPR. There is no doubt, however, that SyRI falls within the scope of the ECHR. All in all, it is a tad strange that the court used the GDPR to elucidate the right to a private life in the ECHR. However, we do not see major problems with that approach.

*Opacity and discrimination risks.* As noted, the court concluded that the SyRI legislation is insufficiently transparent.[76] We agree. It is disturbing how little the government explained about the SyRI system, even to the court in this case.

The court also warned of the risk of discrimination brought by SyRI. Indeed, large-scale data processing for fraud detection creates a risk of discriminating against certain groups (see, from a US perspective, (Eubanks, 2018). Particularly now that we are entering a worldwide recession (triggered by the COVID-19 virus), many governments may be tempted to use digital means to tackle welfare fraud. Such digital fraud detection often brings risks of discrimination. It is innovative that the court connected non-discrimination norms to the right to a private life, as protected under the European Convention on Human Rights.

*The GDPR's provision on automated decision-making.* As noted, the court also discussed Article 22 GDPR, the provision on automated decision-making.[77] The SyRI judgment is the first judgment in the Netherlands that discusses Article 22 GDPR. As an aside: in an unrelated case from 2021, a Dutch court recognised, for the first time in Europe, a right to an explanation regarding an automated decision, based on the GDPR (Gellert et al. 2021).

## Weak points of the judgment

There are two slips of the pen in the judgment concerning, first, the overlap between data protection and privacy; and second, the level of protection the ECHR offers.

---

74. NJCM et al. v The Dutch State (2020) The Hague District Court ECLI: NL: RBDHA:2020:1878 (*SyRI*). Available at: https://uitspraken.rechtspraak.nl/inziendocument?id=ECLI:NL:RBDHA:2020:1878 [6.38]. See also the subpoena by NCJM et al (in Dutch), https://bijvoorbaatverdacht.nl/wp-content/uploads/2019/04/dagvaarding-bodemprocedure-syri-27-maart-2018.pdf.
75. See the Police Directive: Directive (EU) 2016/680 of the European Parliament and of the Council of 27 April 2016 on the protection of natural persons with regard to the processing of personal data by competent authorities for the purposes of the prevention, investigation, detection or prosecution of criminal offences or the execution of criminal penalties, and on the free movement of such data, and repealing Council Framework Decision 2008/977/JHA
76. NJCM et al. v The Dutch State (2020) The Hague District Court ECLI: NL: RBDHA:2020:1878 (*SyRI*). Available at: https://uitspraken.rechtspraak.nl/inziendocument?id=ECLI:NL:RBDHA:2020:1878 [6.86].
77. NJCM et al. v The Dutch State (2020) The Hague District Court ECLI: NL: RBDHA:2020:1878 (*SyRI*). Available at: https://uitspraken.rechtspraak.nl/inziendocument?id=ECLI:NL:RBDHA:2020:1878 [6.35-6.36,6.55-6.60].

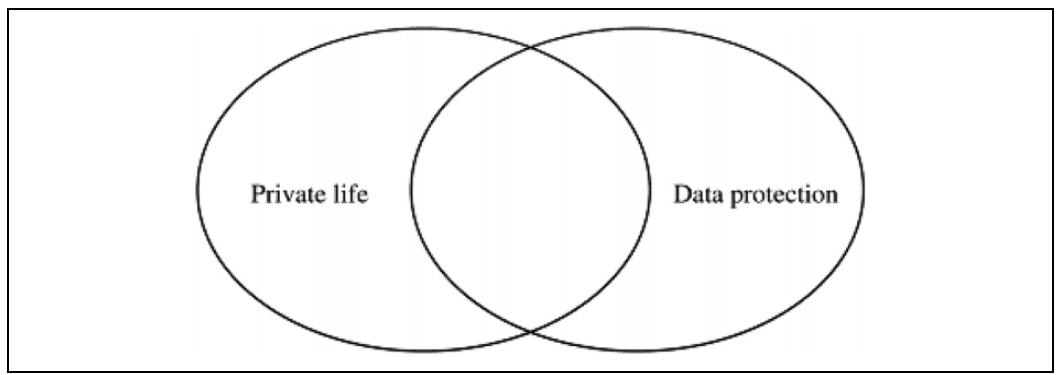

**Figure 2.** The scope of the rights to private life and to the protection of personal data[83].

First, the court spoke of 'the right to respect for private life, which *includes* the protection of personal data' (our emphasis).[78] That is incorrect. The right to a private life (as protected in the ECHR and the Charter of Fundamental Rights of the European Union)[79] and the right to the protection of personal data (a separate right in the EU Charter)[80] partly overlap. But the right to private life does not 'include' the right to the protection of personal data. Some situations fall within the scope of the right to the protection of personal data, while they do not fall within the scope of the right to a private life.

For example, the right to the protection of personal data applies as soon as personal data are processed, even if there is no interference with the right to private a life.[81] To illustrate, the right to the protection of personal data applies to a phone book, because that includes personal data. But being listed in a phone book does not necessarily interfere with the right to a private life.

Meanwhile, some practices violate the right to a private life, while they remain outside the scope of data protection law. For instance, a stalker who follows a victim in the street may violate the victim's right to a private life. However, data protection law only applies if the stalker collects or otherwise processes personal data.[82] The relation between the right to the protection of personal data and the right to a private life is illustrated by Figure 2.

Presumably, the judgment's incorrect phrase was merely a mistake. It is true that the European Court of Human Rights has read several data protection principles into the right to a private life as enshrined in the ECHR. That court had to be so creative in cases regarding personal data processing because the Convention, which dates back to 1950, does not include a separate right to the protection of personal data (De Hert and Gutwirth, 2009: 24–25).

---

78. NJCM et al. v The Dutch State (2020) The Hague District Court ECLI: NL: RBDHA:2020:1878 (*SyRI*). Available at: https://uitspraken.rechtspraak.nl/inziendocument?id=ECLI:NL:RBDHA:2020:1878 [6.6].
79. European Convention on Human Rights Art. 8. Charter of Fundamental Rights of the European Union Art. 7.
80. Charter of Fundamental Rights of the European Union Art. 8.
81. General Data Protection Regulation Art. 2(1). The right to a private life is only mentioned once in the GDPR (recital 4) and 'privacy' is mentioned twice (both in footnote 18).
82. See General Data Protection Regulation Art. 2(1).
83. The figure is taken from Zuiderveen Borgesius, 2015.

The judgment contains a second mistake. The court suggested that the minimum level of protection of the right to privacy in the ECHR is at least as high as the level of protection provided under the GDPR and the Charter of Fundamental Rights of the European Union.[84] That is the wrong way around. The ECHR offers less protection to personal data than the GDPR and the Charter. The European Court of Human Rights has not extended the protection of Article 8 of the ECHR to all personal data. In other words, certain data processing activities do not infringe upon the right to a private life, according to that court.[85] Only if personal data processing concerns data regarding people's private lives, or if data processing is extensive, is the court likely to find that the right to a private life (Article 8 ECHR) is affected. Again, however, this concerns a detail in the SyRI judgment, which is unlikely to cause problems in practice.

Data protection law has developed into a quite technical and specialised field of law. A decade ago, courts sometimes struggled when applying data protection law. In this judgment, however, apart from a few unfortunately phrased sentences, the court's reasoning was solid.

## After the judgment

Soon after the judgment, the Dutch government said that it would not appeal the ruling.[86] Therefore, the judgment is final. What are the implications of the Dutch SyRI judgment for fraud detection and automated decision-making?

The judgment has had direct consequences. SyRI will no longer be used.[87] It is arguably not a big problem for the Dutch state to stop using SyRI, because it appears that SyRI was never very successful in uncovering fraud. (Meuwese, 2020: 209–212)

From a formal legal perspective, the judgment has few implications for other state-run fraud detection systems. The judgment was not about fraud detection of automated decision-making in general, but only about the SyRI system. It is therefore unclear if the rules of the SyRI judgment apply to other fraud-detection systems. In addition, it is unclear to what extent the SyRI system used automated decision-making and it is unknown what exact technology SyRI used (e.g. machine learning, or artificial intelligence). Therefore, the judgment has few legal implications for automated decision-making. One of the main reasons why the court declared SyRI invalid was the lack of transparency (Jansen and Reijneveld, 2020: 371–405). Other fraud detection systems might survive a legal challenge, if the state is more transparent.

Nevertheless, the judgment does emphasise that the government must be transparent in respect of fraud detection systems. While the direct legal effects of the SyRI judgment are limited, it may have indirect effects. Some Members of Parliament said that, in retrospect, they should not have adopted the SyRI legislation without debate.[88]

---

84. NJCM et al. v The Dutch State (2020) The Hague District Court ECLI: NL: RBDHA:2020:1878 (*SyRI*). Available at: https://uitspraken.rechtspraak.nl/inziendocument?id=ECLI:NL:RBDHA:2020:1878 [6.41].
85. After an analysis of the case law of the European Court of Human Rights, De Hert and Gutwirth conclude that 'the Court made a distinction between personal data that fall within the scope of Article 8 ECHR and personal data that do not' (De Hert and Gutwirth, 2009), section 5.1, p. 24-25.
86. Letter of 23 April 2020 to the President of the House of Representatives by the State Secretary for Social Affairs and Employment, Tamara van Ark, on a court judgment regarding SyRI, https://www.rijksoverheid.nl/documenten/kamerstukken/2020/04/23/kamerbrief-naar-aanleiding-van-vonnis-rechter-inzake-syri.
87. Letter of 23 April 2020 to the President of the House of Representatives by the State Secretary for Social Affairs and Employment, Tamara van Ark, on a court judgment regarding SyRI, https://www.rijksoverheid.nl/documenten/kamerstukken/2020/04/23/kamerbrief-naar-aanleiding-van-vonnis-rechter-inzake-syri.

In addition, the SyRI case triggered public debate. The case was extensively discussed in Dutch media and also on national television.[89] Even foreign media discussed the SyRI judgment.[90] Perhaps in future politicians will pay more attention to human rights risks when drafting new laws to tackle welfare fraud. However, it is too early to assess whether it will have such an effect.

## Conclusion

In this article we analysed the SyRI judgment and its implications. States worldwide are turning to technology to make the welfare state more efficient and to mitigate welfare fraud. In the Netherlands, the state used SyRI, a system to combine personal data from different sources and to uncover fraud. A Dutch court declared the SyRI system to be unlawful, because SyRI breached several data protection principles and the right to privacy.

The SyRI system did not strike a fair balance between fraud detection and privacy, according to the court. Most importantly, the Dutch government and the SyRI legislation did not offer sufficient transparency regarding SyRI's use of personal data. We showed that the immediate effects of the judgment are limited. The judgment does not say much about fraud detection and automated decision-making in general. A court might approve a similar system if the government ensures greater transparency.

Still, the SyRI judgment is important. The judgment reminds policymakers that fraud detection must happen in a way that respects data protection principles and human rights such as the right to privacy. The judgment also confirms the importance of transparency about how personal data are used.

### Author contribution

Both authors contributed equally to the paper. The authors would like to thank Maranka Wieringa and Minke Reijneveld for their useful suggestions.

### Declaration of conflicting interests

The author(s) declared no potential conflicts of interest with respect to the research, authorship, and/or publication of this article.

### Funding

The author(s) received no financial support for the research, authorship, and/or publication of this article.

---

88. See e.g. Financieel Dagblad, https://fd.nl/weekend/1367599/scheidend-kamerlid-kees-verhoeven-d66-we-dreigen-controle-over-ons-leven-te-verliezen.
89. See: Wieringa M; Van Schie G; Van de Vinne M (2020). De discussie omtrent SyRI moet over meer dan alleen privacy gaan (The discussion about SyRI should not focus only on privacy). Available at: https://ibestuur.nl/podium/de-discussie-omtrent-syri-moet-over-meer-dan-alleen-privacy-gaan.
90. See e.g. Simonite T (2020) Europe Limits Government by Algorithm. The US, Not So Much. Available at: https://www.wired.com/story/europe-limits-government-algorithm-us-not-much/; Henley J and Booth R (2020) Welfare surveillance system violates human rights, Dutch court rules. Available at: https://www.theguardian.com/technology/2020/feb/05/welfare-surveillance-system-violates-human-rights-dutch-court-rules.

## ORCID iD

Frederik Zuiderveen Borgesius  https://orcid.org/0000-0001-5803-827X

## References

Bekker, S (2021) Fundamental Rights in Digital Welfare States: The Case of SyRI in the Netherlands. In: Spijkers, O, Werner, WG and Wessel, RA (eds) *Netherlands Yearbook of International Law 2019*. Netherlands Yearbook of International Law. The Hague: T.M.C. Asser Press, pp. 289–307. DOI: 10.1007/978-94-6265-403-7_24.

Besselink, LFM (2012) Constitutional Adjudication in the Era of Globalization: The Netherlands in Comparative Perspective. *European Public Law*. Alphen aan den Rijn, The Netherlands: Kluwer Law International: 231–245. Available at: http://www.kluwerlawonline.com/api/Product/CitationPDFURL?file=Journals\EURO\EURO2012013.pdf.

De Hert, P and Gutwirth, S (2009) Data Protection in the Case Law of Strasbourg and Luxemburg: Constitutionalisation in Action. In: Serge, Gutwirth, Poullet, Y, De Hert, Paul, et al. (eds) *Reinventing Data Protection?* Dordrecht: Springer Netherlands, pp. 3–44. DOI: 10.1007/978-1-4020-9498-9_1.

Dommering, E (2020) Annotatie bij Rb. Den Haag 5 februari 2020 (NJCM c.s./Staat der Nederlanden - SyRI-wetgeving). *Nederlandse Jurisprudentie* 45: 6792–6795. Available at: https://www.ivir.nl/publicaties/download/Annotatie_NJ_2020_386.pdf.

Dutch Data Protection Authority [College Bescherming Persoonsgegevens] (2014) *Advies conceptbesluit SyRI*. z2013-00969, 18 February. Available at: https://autoriteitpersoonsgegevens.nl/sites/default/files/atoms/files/z2013-00969.pdf.

Eubanks, V (2018) *Automating Inequality: How High-Tech Tools Profile, Police, and Punish the Poor*. St. Martin's Press.

Gantchev, V (2019) Data protection in the age of welfare conditionality: Respect for basic rights or a race to the bottom? *European Journal of Social Security* 21(1): 3–22. DOI: 10.1177/1388262719838109.

Gantchev, V (2021) SyRI en zijn rechtsopvolger de WGS: Oude wijn in nieuwe zakken [Syri and its legal successor the WGS: old wine in new bottles]. *Tijdschrift Recht en Arbeid* 4.

Gellert, R, Van Bekkum, M and Zuiderveen Borgesius, FJ (2021) The Ola & Uber judgments: for the first time a court recognises a GDPR right to an explanation for algorithmic decision-making https://www.sectorplandls.nl/wordpress/blog/the-ola-uber-judgments/

González Fuster, G (2014) *The Emergence of Personal Data Protection as a Fundamental Right of the EU*. Law, Governance and Technology Series. Cham: Springer International Publishing. DOI: 10.1007/978-3-319-05023-2.

Jansen, RHT and Reijneveld, MD (2020) SyRI-wetgeving in strijd met het EVRM' [SyRI legislation breaches the ECHR]. *Jurisprudentie Bestuursrecht* 4: 371–405.

*Letter of the State Secretary* (2017) 32 761, 122, Kamerstukken II 2017/18.

Mendoza, I and Bygrave, LA (2017) The Right Not to be Subject to Automated Decisions Based on Profiling. In: Synodinou, T-E, Jougleux, P, Markou, C, et al. (eds) *EU Internet Law*. Cham: Springer International Publishing, pp. 77–98. DOI: 10.1007/978-3-319-64955-9_4.

Meuwese, A (2020) Regulating algorithmic decision-making one case at the time. *A note on the Dutch 'SyRI' judgment* 1(1): 209–211.

Parlementaire ondervragingscommissie Kinderopvangtoeslag (2021) *Ongekend onrecht, verslag van de parlementaire ondervraging [Parliament inquiry committee regarding childcare allowance, Unprecedented injustice. Report on the inquiry]*. Report. Tweede Kamer 35 510, nr. 2. Available at: https://www.tweedekamer.nl/sites/default/files/atoms/files/20201217_eindverslag_parlementaire_ondervragingscommissie_kinderopvangtoeslag.pdf.

Raad van State [Council of State] (2014) *Advies Raad van State inzake het ontwerpbesluit houdende regels voor fraudeaanpak door gegevensuitwisselingen en het effectief gebruik van binnen de overheid bekend zijnde gegevens (Besluit SyRI)*. Staatscourant 2014, 26306, 14 August. Available at: https://zoek.officielebekendmakingen.nl/stcrt-2014-26306.html.

U.N. Human Rights Council (2019) *Visit to the United Kingdom of Great Britain and Northern Ireland*. A/HRC/41/39/Add.1, Report of the Special Rapporteur on extreme poverty and human rights, 23 April. Available at: https://ap.ohchr.org/documents/dpage_e.aspx?si=A/HRC/41/39/Add.1.

Wieringa, MA (2021) 'Hey SyRI, tell me about algorithmic accountability: The lessons we can learn from an existing algorithmic system' in *Approaching algorithmic account(-)ability: developing tools to foster formalized and practical transparency in municipal data projects* [Doctoral dissertation in preparation, Utrecht University, Utrecht].

Zuiderveen Borgesius, FJ (2015) *Improving privacy protection in the area of behavioural targeting*. Kluwer Law International.

Zuiderveen Borgesius, FJ (2016). Singling out people without knowing their names – Behavioural targeting, pseudonymous data, and the new Data Protection Regulation. *Computer Law & Security Review*, 32(2), 256–271.